\documentclass{sig-alternate}
\usepackage[show]{chato-notes} 
\usepackage{url}
\usepackage{amssymb}
\usepackage{times}
\usepackage{mathptmx}
\usepackage{graphicx}
\usepackage{multirow}
\usepackage[tight,footnotesize]{subfigure}

\begin{document}

\permission{Submitted for confidential review.}


\title{The Mesh of Civilizations and International Email Flows\titlenote{Part of this work was done while the first author was visiting Yahoo!\ Research Barcelona under the Yahoo!\ internship program.}}
\subtitle{\color{red} No material in this paper may be cited or published in whole \\ or in part without prior written permission of the authors.}
\conferenceinfo{WebSci'13,} {May 2--4, 2013, Paris, France.} 
\CopyrightYear{2013} 
\crdata{978-1-4503-1869-3/13/02} 
\clubpenalty=10000 
\widowpenalty = 10000

\numberofauthors{3}
\author{
\alignauthor
	Bogdan State\\
	\affaddr{Stanford University}\\
	\email{bstate@stanford.edu} 
\alignauthor
	Patrick Park\\
	\affaddr{Cornell University}\\
	\email{pp286@cornell.edu} 
	\alignauthor
	Ingmar Weber\\
	\affaddr{Qatar Comp. Res. Inst.} \\
	\email{ingmarweber@acm.org} 
	\and
\alignauthor
	Yelena Mejova\\
	\affaddr{Yahoo!\ Research} \\
	\email{ymejova@yahoo-inc.com}
	\alignauthor
		Michael Macy\\
	\affaddr{Cornell University}\\
	\email{m.macy@cornell.edu} 
}

\maketitle

\begin{abstract}
In \textit{The Clash of Civilizations}, Samuel Huntington argued that the primary axis of global conflict was no longer ideological or economic but cultural and religious, and that this division would characterize the ``battle lines of the future.'' In contrast to the "top down" approach in previous research focused on the relations among nation states, we focused on the flows of interpersonal communication as a bottom-up view of international alignments. To that end, we mapped the locations of the world's countries in global email networks to see if we could detect cultural fault lines. Using IP-geolocation on a worldwide anonymized dataset obtained from a large Internet company, we constructed a global email network. In computing email flows we employ a novel rescaling procedure to account for differences due to uneven adoption of a particular Internet service across the world. Our analysis shows that email flows are consistent with Huntington's thesis. In addition to location in Huntington's 
``civilizations,'' our results also attest to the importance of both cultural and economic factors in the patterning of inter-country communication ties.
\end{abstract}
\keywords{social networks, email, international networks}

\section{Introduction}

Are the world's countries re-aligning as the buildup to a global culture war? Most research has examined international global alignments from the top down, based on the relations among nation states. Rather than examining the relations among states, we take a bottom-up view by examining the flows of email between countries, to map global patterns of cross-national integration and division based on the structure of interpersonal social ties among the populations of the world's countries. 

Our study extends this line of research on spatial and geographic patterns by examining economic, demographic, and cultural correlates of international communication densities. We estimate these densities using an anonymized collection of email exchanges numbering in order-$10^7$ of users. 

To account for the uneven distribution of the email service's market share, we develop a novel procedure for rescaling the communication densities. To do so we regress the realized between-country density on the number of users in the sample. Using this regression we then predict the most likely value of the tie count between the full populations of two countries, rather than just between their email users.

Using this network of cross-country affinities, we investigate the covariance of a battery of cultural measures with inter-country flows of interpersonal email communication. Following Huntington, we code countries based on location in one of the ``civilizations'' that he demarcates, using data derived by Russett, Oneal and Cox \cite{Russettetal:2000, Neumayer:2011}. The cultural variables we consider include Hofstede's \cite{Hofstede:1980} Power-Distance (PDI), Individualism (IDV), Masculinity (MAS) and Uncertainty Avoidance (UAI), and Bjornskov's generalized trust index \cite{Bjornskov:2007}. We also examine the role of economic and political factors, including the GDP and membership in the European Economic Area. Lastly, we included demographic measures of population size and distance. 

Our analysis reveals the existence of a large, positive statistically-significant effect of common civilizational membership on between-country communication density. This result provides evidence towards a division of the world into civilizational blocks following Huntington's theory. We find that Huntington's partitioning of countries has about the same level of agreement with the results of community detection algorithms as such algorithms have with one another. We also uncover effects due to economic inequality, as posited by World Systems Theory, as well as a robust effect due to Hofstede's Uncertaint Avoidance measure. 



\section{Related Work}

\textbf{Geo-social datasets}. 
The emergence and growth of internet platforms has enabled researchers to study very large networks recorded through Internet communication\cite{Kleinberg:2008}. The scaling properties of large networks -- social and otherwise -- emerged from analyses made possible by web data\cite{Barabasi:2009, WattsDoddsNewman:2002}. Methodological advances have opened up new research questions related to community detection in large social networks \cite{ClausetNewmanMoore:2004, AhnBagrowLehman:2010} or to the distribution of shortest paths \cite{Backstrometal:2011}.

New technologies have also facilitated the collection and study of large amounts of geographic data, enabling previously-unthinkable studies of the spatial properties of social interaction. Mobile phone geo-location has been used to infer friendship structure \cite{EaglePentlandLazer:2009}, and even to measure country-level social network \cite{EagleMacyClaxton:2010, BlumenstockGillickEagle:2010}. Location-based websites such as Foursquare or Gowalla have also proven to be useful data sources for the study of geo-social interaction. Gowalla data has been used to study the relation between social networks and mobility, uncovering a high level of social determination of the local mobility patterns of individuals \cite{SamalaKingsford:2012}. Geo-social data, web- or cellphone-based has already proven its usefulness in applications as diverse as epidemiology \cite{SamalaKingsford:2012, Tatemetal:2012}, public transportation \cite{MachadoJoseMoreira:2012}, estimation of migration rates 
\cite{ZagheniWeber:2012} or event recommendation~\cite{SklarShawHogue:2012}. 

A bottom-up approach to measuring trans-national social networks has become feasible only recently. Though more accurate methods for the collection of geographic information have been developed recently, data coverage has hindered their use in a global setting, as cellphone networks and location-based web services typically cover only one or a few countries. To our knowledge, the earliest study that tackled the issue of global online transnational patterns was conducted by Leskovec and Horvitz (2008), whose measurements provided evidence of large communication flows between countries with colonial pasts (e.g.\ Portugal and Brazil), countries that are close geographically (e.g.\ France and Belgium), or countries connected by histories of large migrations (e.g.\ Germany and Turkey) \cite{LeskovecHorvitz:2008}. In a study of the CouchSurfing international hospitality network Lauterbach et al.\ (2009) described the transnational web of hospitality exchange ties connecting the members of the organization \cite{
Lauterbachetal:2009}. Both studies addressed the issue of transnational networks only tangentially, however, relying only on self-reported location information. Our work addresses this gap in the literature by attempting the first study focused primarily on the international structure of worldwide social networks. 

The problem of studying inter-national networks is compounded by the comparative scarcity of between-country ties, relative to ties formed within the same country, as a number of recent studies have noted. Scelatto et al.\ (2011) note that most social ties are separated geographically by small distances \cite{Scelattoetal:2011}. This finding is reproduced by Takhteyev et al.\ (2012) in a recent study of Twitter networks that shows that most connections occur within the same country \cite{Takhteyevetal:2012}. The study of a large, global network is therefore necessary to obtain a clear enough view of the global transnational communication networks. 

\textbf{Cross-country Affinity}.
For most of the postwar period, research on international alignments was informed by World Systems Theory, an approach that emphasized the influence of economic and political factors. Broadly speaking, World Systems Theory posits the existence of a hierarchical structure in international relations, in which a number of core countries engage in the exploitation of peripheral states (often corresponding to former colonial empires). Simply put, international alignments are believed to be structured by relations of global economic inequality \cite{ChaseDunnGrimes:1995}. 

In the early 1990's, Samuel Huntington called this economic model into question. In \textit{The Clash of Civilizations}, Samuel Huntington argued that:
\begin{quote}
``the fundamental source of conflict in this new world will not be primarily ideological or primarily economic. 
The great divisions among humankind and the dominating source of conflict will be cultural. Nation states will remain the most powerful actors in world affairs, but the principal conflicts of global politics will occur between nations and groups of different civilizations. The clash of civilizations will dominate global politics. The fault lines between civilizations will be the battle lines of the future.'' 
\cite{Huntington:1993}
\end{quote}

Other scholars have also pointed to cultural correlates of economically structured international alignments. Banfield (1958) argued that Southern and Northern Italy fundamentally differ in cultural norms that account for striking differences in economic development \cite{Banfield:1958}. This idea received further elaboration in Putnam, Leonardi and Nannetti's (1994) work on Italian regionalization, which introduced the idea of differences in the structure of social networks between different societies \cite{Putnam:1994}. More recently, the concept of generalized trust -- the extent to which individuals can trust others -- has gained credence as a fundamental characteristic of social interaction in different societies, high levels being associated with economic development and efficient institutions \cite{Fukuyama:1995, Bjornskov:2009, Nannestad:2008}.

Hofstede's seminal work in the 1980's also identified differences in cultural values on a global scale. Hofstede \cite{Hofstede:1980} designed a survey administered to IBM employees from 55 nations, probing a wide range of cultural values related to authority relations, the relationship between individual and society, gender roles, and social and environmental uncertainties. From the IBM study Hofstede derived a number of cultural dimensions, including the four measures used in our study for which data are widely-available: the power-distance index (PDI), individualism-collectivism (IDV), masculinity-femininity (MAS), and the uncertainty avoidance index (UAI). Power-distance (PDI) measures the extent to which individuals accept unequal power distribution in their social relationships. Individualist (IDV) societies are defined as societies where ties among individuals are loose and members perceive themselves as independent self-reliant entities endowed with freedom and responsibility. Masculinity (MAS) 
measures a society's level of distinction in gender roles, where men are expected to be ``assertive, tough, and focused on material success'' while women are expected to be ``modest, tender, and concerned with quality of life'' \cite{HofstedeMinkov:2010}. Hofstede characterizes a country as ``masculine'' when these gender roles are distinct and ``feminine'' when they overlap. Uncertainty avoidance (UAI) refers to the extent to which a society is intolerant and threatened by uncertain situations.

\section{Data and Methods}\label{sec:data}

Our work is based on the aggregate analysis of a communication graph composed of a sample of order-$10^7$ anonymized users of Yahoo!\ email,
observed over a period of 6 months in 2012. An edge was considered to exist between a pair of users whenever the two users exchanged at least one email message  in each direction, during the observation window. Only users with a minimum degree of one were included in our analysis, and our study included only those users who were not identified as spammers, and who had given consent for their email data to be studied. In order to create the communication graph, our study processed only the email header fields indicating the sender and recipient's email address.

\subsection{Inferring Location}\label{sec:location}

We identified a user's country from two independent sources: the user's self-identified country, as recorded in the user profile database and the IP geo-location. Users often expedite entry of their location by selecting one of the first countries on the drop down menu (e.g.\ Afghanistan), and users do not always update their profile when moving to a different country. We therefore combined self-reported address with information derived from IP geolocation. Our analysis used a similar protocol to that implemented by State, Weber and Zagheni in their study of international migrations \cite{StateWeberZagheni:2013}. We used the MaxMind GeoCityLite database\footnote{\url{http://dev.maxmind.com/geoip/geolite}} to extract coarse-grained, city-level geographic information associated with each IP address from which a user logged in during an observation window of about one year. We divided the data into spells in a similar fashion to the protocol used in \cite{StateWeberZagheni:2013}.\footnote{As our analysis was 
concerned only with the modal country in which a user 
was observed, we used slightly less stringent assumptions however. We allowed valid international transitions to have a maximum implicit speed of 1000, rather than 150 km/h. Additionally, we considered as validly identified through geo-location those users for whom the cumulative duration of valid spells exceeded a threshold of 90 days, rather than 300.} We considered the geolocated country of residence to be the modal country from which the user was observed to log in, as per our protocol. Our analysis was further restricted to those users for whom both self-reported and geo-located country of residence coincided. We additionally imposed minimum thresholds on the number of valid users in a country that could be included in the study, discarding countries having too few users in our dataset. 

Next, we collapsed the data to a matrix of $c \times c$ countries, each cell indicating the observed tie density between two countries corresponding to the number of observed reciprocal email ties between individuals in the two countries divided by the total possible number of ties, given the total number of individuals observed in each country.

\subsection{Rescaling Procedure}\label{sec:rescaling}

Finally, we developed a rescaling procedure to address significant potential biases resulting from differences in market penetration and Internet use. Our aim is to estimate tie densities between two countries, but we wish to factor out the effects of uneven data coverage. For instance, our counts could be off by several orders of magnitude between two countries where (due to low Internet use or low market share) we observe a fraction of a percent of the population, as compared with countries where our observations concern ties between 20\% of each of the two populations. To properly rescale the communication densities we wish to produce an accurate estimate of the total number of social ties between countries $i$ and $j$ ($T_{i, j}$) and ties within the same country, when $i=j$. $T_{i,j}$ is bounded between 0 and $T^{\mathrm{max}}_{i,j} =N_i N_j$ if $i \neq j$, and between $0$ and $T^{\mathrm{max}}_{i}$ = $N_i(N_i - 1)/2$ if $i = j$. Let $t_{i,j} = T_{i,j} / T_{i,j}^{\textrm{max}}$ the proportion of ties 
observed between countries $i$ and $j$. Furthermore, let $c_i$ be the fraction of country $i$'s adult population $P_i$ that is currently represented in the data. Thus $N_i = c_iP_i$ and $N_j = c_jP_j$.

Assume we move from observing fraction $c_i$ of a country's population to collecting data about all individuals in the country. $N_i$ would then increase by a factor of $1/c_i$ to equal $P_i$. The new maximum count will be updated accordingly. If $i \neq j$, then: 

$$T_{i,j}^{\textrm{max'}} = (c_ic_j)^{-1} T_{i,j}^{\textrm{max}}$$

For sufficiently large $N_i$'s the same relation holds as an approximation for the case when $i = j$. At first blush, the normalization procedure would attempt to preserve the density $t_{i,j}$ constant, and thus use the formula $T_{i,j}' = (c_ic_j)^{-1} T_{i,j}$ to rescale the observed tie-count by the same factor by which the possible tie-count increases. This approach would be misguided, however. As social graphs grow by orders of magnitude, density does not stay constant. The simplest way to justify this claim employs Dunbar's number \cite{Dunbar:2010}, the empirically-verified limit -- often quoted as 150 alters -- of social ties a person can maintain at one point in time. Were a growing graph to maintain constant density, the mean degree would have to grow linearly with the number of nodes, eventually overshooting Dunbar's number.

\begin{figure}
\includegraphics[width=\columnwidth]{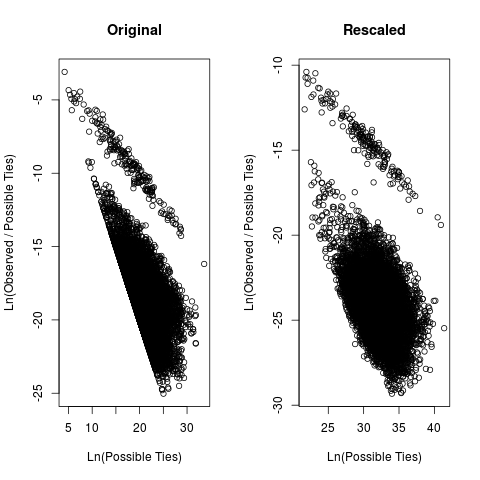} 
\vspace{-2em}
\caption{Email communication Log-densities, before and after rescaling.} 
\label{fig:density}
\vspace{-2em}
\end{figure}

This observation is verified in practice by the graph presented in Figure \ref{fig:density}. 
The graph plots along logarithmic axes the tie density $t_{i,j}$ between a pair of countries $(i,j)$ graphed and the maximum number of ties between the countries. The graph shows a linearly decreasing bound of maximum density, obtained for the case when we are considering ties within the same country. Rescaling the tie counts requires attention not only to the absolute vertex count of the observed graph ($N_i$), but also to the share of a country's population contained in the graph ($c_i$). Web-based services such as Yahoo!\ do not grow randomly; rather, the graph expands through the social networks of current users. This process resembles multi-seeded snowball sampling, starting from a few unrelated individuals (the early adopters), and expanding to an increasing number of the current users' social contacts who decide to join the network. When in one country only a small proportion of users is included, it is quite 
plausible that idiosyncratic variances in likelihood of having social connections with certain other countries would be correlated between individuals in the sample, who are likely to be clustered together. Thus, the positive ``signal'' observed at low levels of market penetration with respect to between-country connections is likely to be overstated in relation to the total possible tie-count, compared to what one would observe at higher levels of market penetration.

We use a log-linear regression model to predict the expected decrease in the density $t_{i,j}$, as a function of the size of each country's userbase, counted both in absolute terms and as a proportion of each country's adult population. The regression results support our assumptions, as the predicted density decreases with the number of users and their share in the population (Table \ref{tbl:rescaling-model}). As Figure \ref{fig:density} shows, the data is organized in two clusters, one of within-country densities, the other of between-country densities. We estimate the model jointly for the two clusters, but allow for within-cluster variations in the effects of each predictor, through the use of interaction effects.

\begin{table}
\parbox{\columnwidth}{\caption{Ordinary Least Squares Regression. Response: Ln Between-Country Ties / Total Possible Ties}\label{tbl:rescaling-model}}
\vspace{-1em}
\begin{center}
\begin{minipage}{\columnwidth}
\begin{tabular*}{\columnwidth}{ l @{\extracolsep{-1em}} r r r}
\hline
\multicolumn{1}{c}{\textbf{Independent Variable}} & 
\multicolumn{1}{c}{\textbf{Coefficient} \hspace{-1em}} &
\multicolumn{1}{c}{\textbf{(S.E.)} \hspace{-1em}} &
\multicolumn{1}{c}{\textbf{T-value} \hspace{-1em}} \\
\hline
\hline
Intercept 						& -10.97$^*$	&  (.26)	& -42.12 \\
Users in Country \#1$^\dagger$ 			&  -0.36$^*$	&  (.01)	& -36.57 \\
Users in Country \#2$^\dagger$ 			&  -0.35$^*$	&  (.01)	& -34.48 \\
$\mathrm{Users}_1 / \mathrm{Pop}_1^\dagger$	 	&  -0.13$^*$	&  (.01)	& -10.21 \\
$\mathrm{Users}_2 / \mathrm{Pop}_2^\dagger$	 	&  -0.15$^*$	&  (.01)	& -12.03 \\
Mean Degree of Country \#1				&  -1.68$^*$	&  (.16)	& -10.91 \\
Mean Degree of Country \#2				&  -1.41$^*$	&  (.18)	& -7.97 \\
Mean Degree of Country \#1 $\times$ \#2		&   1.45$^*$	&  (.17)	& -8.02 \\
Same country						&  10.52$^*$	&  (1.04)	& 10.15 \\
$\dots \times $ Users $^\dagger$			&  -0.17$^*$	&  (.07)	& -2.62 \\
$\dots \times $ Users / Population $^\dagger$		&   0.41$^*$	&  (.08)	&  4.20 \\
$\dots \times $ Mean Degree				&   1.34$^*$	&  (.40)	&  3.08 \\
\hline
\end{tabular*}
\footnotesize
{\it Source:} $\dagger$ - Transformed by taking natural logarithm. Sample size: 9,144 . *: p $<$ .001, +: p$<$.01. Two-tailed tests. Adjusted $R^2$: .58.
\vspace{-2em}
\end{minipage}
\end{center}
\end{table}

The $R^2$ coefficient of .58 shows that this very simple regression model explains nearly 60 percent of the variance in changes in the observed density, with international dynamics expect to account for the remainder of the variance. The model allows an important adjustment: We can derive the expected density $t_{i,j}'$ by refitting the model using the country's population as the user base (and thus assuming a hypothetical sample of 100\% of a country's adults). The estimate $t_{i,j}'$, represents the model's best guess as to what the density of ties between a particular pair of countries should be, in the hypothetical scenario of a network census being available for both countries. Accordingly,
\begin{equation}
t_{i,j}' = \frac{T_{i,j}'}{T_{i,j}^{\textrm{max}'}} = \frac{T_{i,j}'}{(c_i c_j)^{-1}T_{i,j}^{\textrm{max}}} = c_ic_j \cdot T_{i,j}' \cdot (T_{i,j}^{\textrm{max}})^{-1}
\label{eqn:density}
\end{equation}
To rescale the ties we can then divide $t_{i,j}'$ through the original density $t_{i,j} = T_{i,j} / T_{i,j}^{\textrm{max}}$. The resulting fraction gives:
$$ \frac{t_{i,j}'}{t_{i,j}} = \frac{c_i c_j T_{i,j}' (T_{i,j}^{\textrm{max}})^{-1}}{T_{i,j} (T_{i,j}^{\textrm{max}})^{-1}} = \frac{c_i c_j T_{i,j}'}{T_{i,j}} $$
From this relation we can extract the following formula for $T_{i,j}'$:\vspace{-5pt}
$$ T_{i,j}' = (c_ic_j)^{-1} \frac{t_{i,j}'}{t_{i,j}} T_{i,j} $$\vspace{-5pt}
In other words, it is now possible to correct the rescaling by multiplying by the ratio $\frac{t_{i,j}'}{t_{i,j}}$, which quantifies how much the density would decrease if the network of all individuals in the two countries were observed. We derive this ratio by dividing the predicted values of both $t_{i,j}'$ \textit{and} $t_{i,j}$.\footnote{We divide through the predicted, and not the observed, density under the current sampling conditions so as not to impose any further assumptions than are necessary on the data. Dividing through the observed density would have imposed a strictly linear relation between the population of countries and the tie count, thus eliminating precisely the variance that will make the object of further study in the paper.}

\section{International Structure in Social Networks}

\begin{figure*}[t]
\begin{center}
\subfigure[Rescaled Counts]{
\includegraphics[width=\columnwidth]{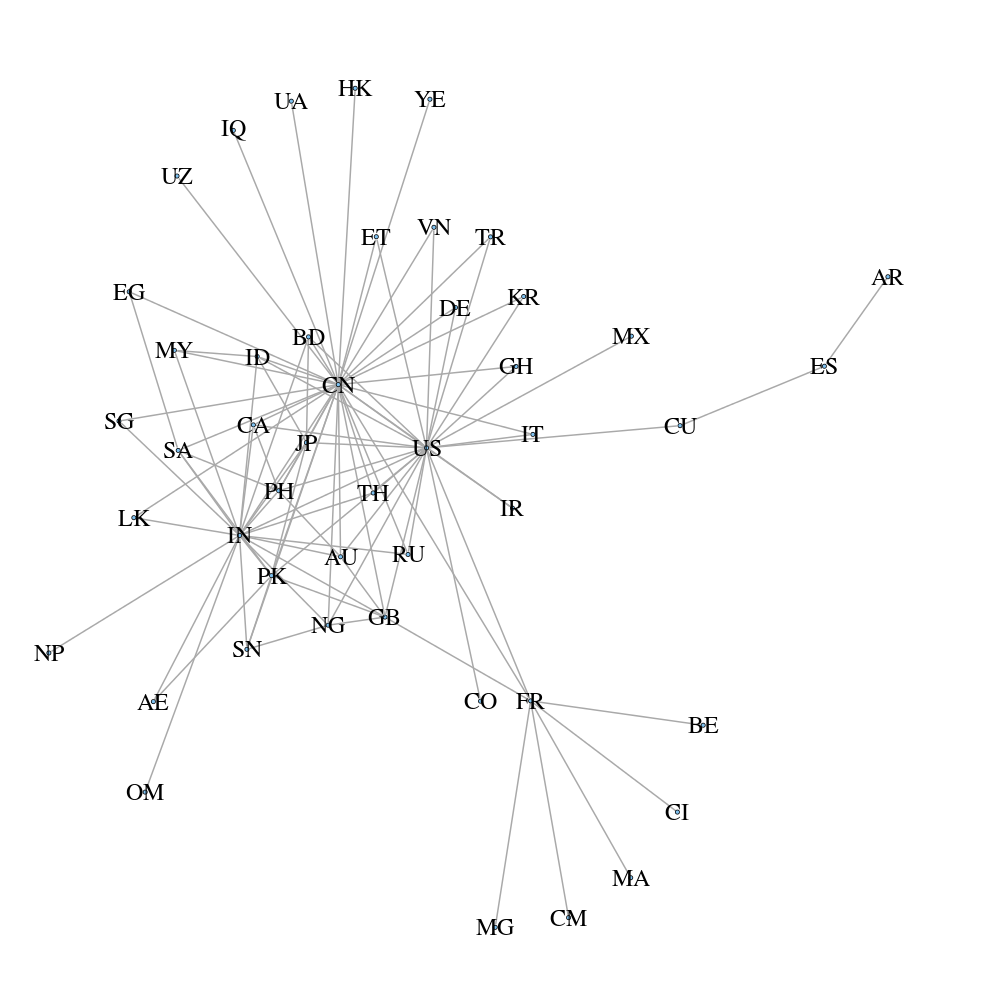}}
\subfigure[Normalized by Population Size]{
\includegraphics[width=\columnwidth]{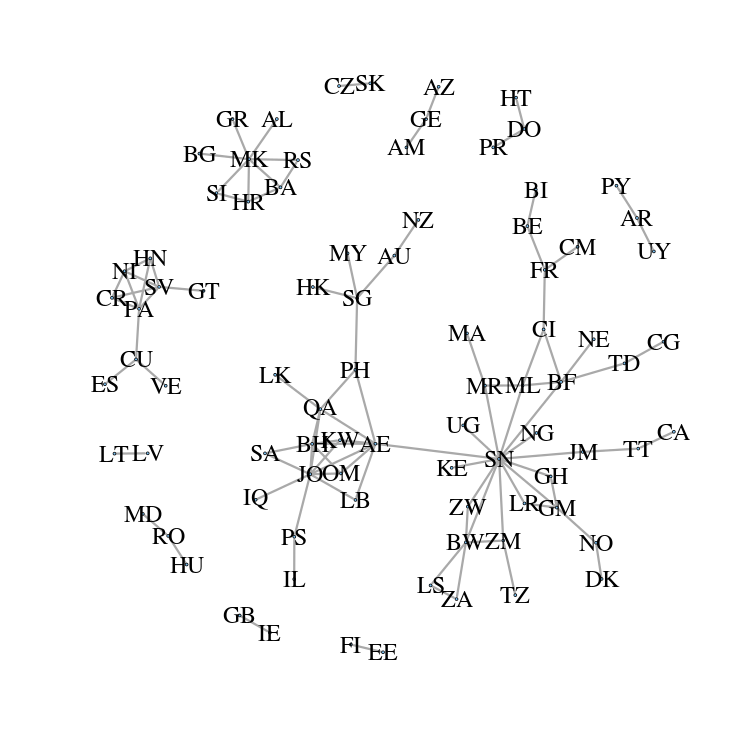}}
\vspace{-1em}
\caption{\label{fig:net-graphs} Top 100 rescaled counts, raw and normalized.}
\vspace{-2em}
\end{center}
\end{figure*}

The pairwise densities estimated through the rescaling procedure can be represented as a weighted network of cross-country connections. Figure \ref{fig:net-graphs}(a) 
 represents the top 100 largest between country ties, in terms of their absolute size. 
 As expected, most of the ties occur between countries with large populations. Though useful in identifying the highest magnitude cross-country ties, this representation communicates little about the deeper structure of the world's social networks. A more useful picture can be obtained by inspecting the top 100 ties between countries, judged by their rescaled densities ($t_{i,j}'$), defined as the ratio between the rescaled raw tie counts and the total number of possible ties (Equation \ref{eqn:density}). Given that for even the smallest countries such values are going to be extremely low, all of our calculations are carried out in log-space to prevent numerical underflow, and to drastically improve model fit.
 

We obtain a graph of 141 countries and 7,246 ties out of 9,870 possible.\footnote{We imposed a threshold for each count: country pairs with too few connections were recorded as having none}
The tie weights are derived from the rescaled logarithm of the communication densities. Because the logarithms are negative, the minimum observed log-density ($t_{\mathrm{min}}'$)  is subtracted from each cell of the adjacency matrix:
$$ w_{i,j} = \mathrm{ln }\;t_{i,j}' - \mathrm{ln }\;t_\mathrm{min}'$$
The resulting edge weights $w_{i,j}$ thus indicate how many times over (in terms of powers of the number $e$) a certain between-country rescaled density $t_{i,j}'$ exceeds the minimum rescaled density $t_\mathrm{min}'$.\footnote{The natural logarithm of the minimum rescaled density is -29.36, corresponding to one expected cross-border tie between individuals in two countries for every 5.6 trillion 
possible. A rescaled density of -18 corresponds to one tie for every 66 million. By subtracting the minimum value of -29.36 from -18 we get instead a value of 11.36, indicating that the observed count is 86,000 times greater than the minimum possible count.} The graph thus indicates a logarithmic measure of affinity between countries, indicating orders of magnitude, rather than absolute counts.  

\begin{figure*}[p]
 \includegraphics[width=\textwidth]{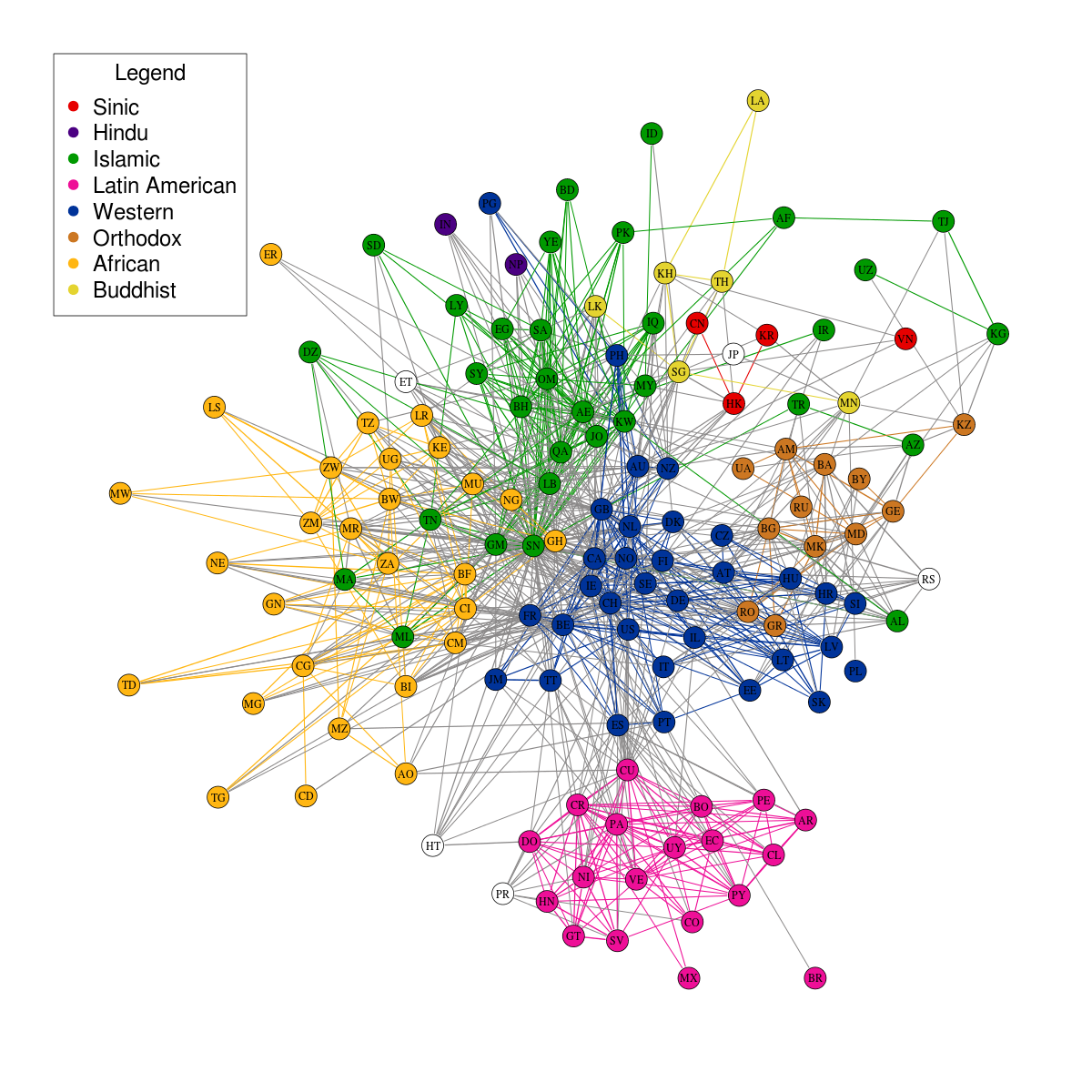}
 \caption{The Mesh of Civilizations}
 \label{fig:mesh}
 \footnotesize
 \textit{Source:} Yahoo!\ email dataset. Rescaled densities. Only top 1,000 densities displayed. Colors indicate Huntingtonian civilization, as collected by \cite{Russettetal:2000} and provided by \cite{Neumayer:2006}. Layout using weighted Fruchterman-Reingold algorithm \cite{FruchtermanReingold}, as implemented in igraph R package \cite{CsardiNepusz:2006}. Layout based on full graph of rescaled communication densities, using monotonic transformation $f(x) = [(x - \textrm{min}(x)) / \textrm{range}(x)]^4$, where $x$ is the natural logarithm of the communication density. Only countries with more than 1m inhabitants as per \cite{WorldBank:2011} included. Observations on Somalia, Myanmar and the Palestinian Territories excluded.
 \label{fig:corr}
\end{figure*}

Figure \ref{fig:mesh} plots the top 1,000 ties observed in the above-described graph, laid out according to the Fruchterman-Reingold algorithm \cite{FruchtermanReingold, CsardiNepusz:2006}, and nodes are colored according to their presumed civilizational membership. Upon visual inspection the graph provides evidence for Huntington's theory. The graph shows clear clusters according to civilizations. The Latin American cluster is most striking, set off from the rest of countries in one region of the graph, with Spain and Portugal -- the former colonial metropolises -- acting as intermediaries between this civilization and the Western civilization, which likewise occupies its own clear region of the graph, with the exception of the Philippines and Papua New Guinea, two countries which can be judged as marginal to the Western block. The Orthodox civilization (ochre) is contiguous with the Western (blue) region of the graph, with Greece and Kazakhstan in-between the Orthodox cluster and the Western and the 
Islamic 
regions, respectively.
The Islamic civilization appears less coherent, with Central Asian, Middle Eastern and North African countries in separate regions, though with some level of contiguity. Sub-Saharan African countries appear torn between two tendencies - to connect within their civilization, or to connect outside the civilization, to Western former colonial powers, or to Middle Eastern countries, with which some Sub-Saharan African countries share religious affinities. 

The visual representation of the graph shows clear hints of a correlation between the labeling of countries according to civilization and the obtained structure of the world's communication network. Indeed, the adjacency matrix obtained by creating a graph of co-civilizational memberships has a product-moment correlation coefficient of .397 with the adjacency matrix of the rescaled communication network.\footnote{Correlation calculated using the gcor function in the SNA R package \cite{Butts:2012, Krackardt:1987}. Adjacency matrix formed by natural logarithm of rescaled communication densities first normalized by subtracting lowest density obtained.} This result's statistical significance is bolstered by a test using the Quadratic Assignment Procedure (QAP) \cite{Krackardt:1987}. Given a certain graph structure (i.e., the communication network) and a certain set of vertex labels (i.e., civilizational membership), QAP computes permutations, thus generating alternative, random partitions of the world's 
countries. No such permutation approaches the obtained correlation coefficient: out of 10,000 random civilizational assignments, the highest obtained correlation coefficient was .059, less than a sixth of the correlation obtained using Huntington's labeling.\footnote{Performed using qaptest procedure in SNA R package~\cite{Butts:2012, Krackardt:1987}~.}

Another observation related to Figure \ref{fig:mesh} 
concerns the central position of the Western civilization compared to the others. To test whether Western countries are truly central to the derived communication graph we compute three measures of centrality, reported in Table \ref{tbl:cent}. Degree centrality indicates the total weighted degree of each country.This measure translates to $\mathrm{deg}_i = \ln (\prod_{j \neq i} t_{i,j}' / t_{\mathrm{min}})$, where $t_\mathrm{min}$ is the minimum observed density.

Western countries have the highest mean degree centrality (1302.4), followed by Sinic (1076.8) and Islamic (1029.6) countries, while the lowest values are recorded for Latin American (904.4) and African (806.6) countries. Eigenvector centrality \cite{Bonacich:1987} indicates the extent to which a country has large rescaled communication densities with other countries that have similarly large densities. Western countries are again at the top of this ranking, with a mean centrality score of .101, followed by Orthodox, Sinic and Islamic countries, with scores of .084, .083 and .081, respectively, while African countries have the lowest score (.064). Betweenness centrality \cite{Freeman:1977} represents an alternative conceptualization of position in the network, expressing the extent to which a country lies on the (weighted) shortest paths between other countries in the graph. Sinic countries have the highest betweenness score (60.79), followed by Hindu (58.46) and Western (58.06) countries, whereas the lowest 
scores register with Orthodox (27.79) and Latin American countries (17.25). 

Seen from the perspective of network analysis, Huntington's effort may be conceived as one at partitioning the world's graph of between-country affinities into a series of communities. A simple question to ask is whether one could do ``better'' than Huntington at partitioning the world's countries into civilizational blocks, based on the structure observed in the communication network. We compared Huntington's assignments to those made by three community detection algorithms for weighted undirected graphs: the Spinglass algorithm \cite{ReichardtBornholdt:2006}, the Walktrap algorithm \cite{PonsLatapy:2005}, as well as the greedy algorithm proposed by Clauset, Newman and Moore \cite{ClausetNewmanMoore:2004}.\footnote{We ran this analysis using the igraph R package \cite{CsardiNepusz:2006}.} Cross-tabulated community assignments are shown in Table \ref{tbl:clustering}. The African, Latin American, Orthodox, Hindu and Sinic civilizations appear to be particularly consistent, all countries in each one of the 
five 
civilizations being assigned to the same community by two of the algorithms. An examination of the Rand index \cite{HubertArabie:1985} reveals that the best agreement occurs between Huntington's assignments and the Walktrap algorithm, the two graph partitions being in agreement about 42.7\% of all pairs of countries being in the same community or not. By comparison, the cross-tabulation of Walktrap and Spinglass had a Rand index of 39.7\%, the same value being 28\% for a cross-tabulation of Walktrap and Greedy.

\setlength{\tabcolsep}{2pt}
\begin{table}[ht]
\begin{center}
\caption{\textbf{Community Detection Results across Civilizations}}
\label{tbl:clustering}
 \begin{tabular}{| l | r r r r | r r r r r r r r | r r r |}
  \hline
\textbf{Civ.} & \multicolumn{4}{| c |}{Spinglass} & \multicolumn{8}{| c |}{Walktrap} & \multicolumn{3}{| c |}{Greedy}\\
\hline
  \hline
 Cross-tab & 1 & 2 & 3 & 4 & 1 & 2 & 3 & 4 & 5 & 6 & 7 & 8 & 1 & 2 & 3 \\ 
  \hline
African &  28 &   0 &   0 &   0 &  17 &   0 &   2 &   0 &   0 &   2 &   2 &   5 &   0 &  28 &   0 \\ 
Buddhist &   0 &   1 &   5 &   0 &   0 &   1 &   3 &   0 &   2 &   0 &   0 &   0 &   1 &   5 &   0 \\ 
Hindu &   0 &   0 &   2 &   0 &   0 &   0 &   0 &   0 &   2 &   0 &   0 &   0 &   0 &   2 &   0 \\ 
Islamic &   7 &   8 &  17 &   0 &   0 &   8 &   2 &   0 &  17 &   0 &   0 &   5 &   8 &  24 &   0 \\ 
Lat. Am. &   0 &   0 &   0 &  19 &   0 &   0 &   0 &  19 &   0 &   0 &   0 &   0 &   0 &   0 &  19 \\ 
Orthodox &   0 &  12 &   0 &   0 &   0 &   8 &   4 &   0 &   0 &   0 &   0 &   0 &  12 &   0 &   0 \\ 
Sinic &   0 &   0 &   4 &   0 &   0 &   0 &   4 &   0 &   0 &   0 &   0 &   0 &   3 &   1 &   0 \\ 
Western &   4 &  17 &   4 &   8 &   0 &   0 &  24 &   0 &   0 &   5 &   0 &   4 &  20 &   8 &   5 \\ 
   \hline
Rand Index 
& \multicolumn{4}{| c |}{0.371}
& \multicolumn{8}{| c |}{0.427}
& \multicolumn{3}{| c |}{0.271} \\
\hline
$\chi^2$ stat. 
& \multicolumn{4}{| c |}{239.84}
& \multicolumn{8}{| c |}{352.14}
& \multicolumn{3}{| c |}{172.88} \\
dF
& \multicolumn{4}{| c |}{21}
& \multicolumn{8}{| c |}{49}
& \multicolumn{3}{| c |}{14} \\
\hline
\end{tabular}
\vspace{-1em}
\end{center}
\end{table}

\begin{table}
 \parbox{\columnwidth}{\caption{Mean Weighted Centrality Scores, by Civilization}}
 \label{tbl:cent}
 \vspace{-1em}
 \begin{center}
  \begin{tabular}{| l  | r r r |}
  \hline
   \multirow{2}{*}{\textbf{Civilization} \hspace{32pt} } &
   \multicolumn{3}{c |}{\textbf{Centrality}} \\
   \cline{2-4} 
   & Degree & Betweenness & Eigenvector \\
   \hline
   African		& 806.6		& 32.06	& 0.064 \\
   Buddhist		& 949.2		& 31.32	& 0.076 \\
   Hindu		& 914.8		& 58.46	& 0.073 \\
   Islamic		& 1029.6	& 33.83	& 0.081 \\
   Latin American	& 904.4		& 17.25	& 0.072 \\
   Orthodox		& 1052.9	& 27.79	& 0.084 \\
   Sinic		& 1075.8	& 60.79	& 0.083 \\
   Western		& 1302.34	& 58.06	& 0.101 \\
   \hline
  \end{tabular} 
  \footnotesize
  \parbox{0.9\columnwidth}{\small
\textit{Source:} Yahoo!\ email dataset. Rescaled densities. Statistics based on adjacency matrix of natural logarithm of rescaled communication densities, transformed by subtracting the minimum observed value. Values calculated using SNA package in R \cite{Butts:2012}.}
 \end{center}
  \vspace{-2em}
\end{table}

\section{Determinants of Inter-Country Connections}

\setlength{\tabcolsep}{2pt}
\begin{table}[t!]
\parbox{\columnwidth}{\caption{Linear mixed-effects model. Response: Log-Density}}
\label{tbl:bet-det}
\begin{center}
\begin{minipage}{\columnwidth}
\vspace{-1em}
\begin{tabular}{ l  r r r r}
\hline
\multicolumn{4}{l}{\textbf{FIXED EFFECTS}}\\
\hline
\multicolumn{1}{c}{\textbf{Indep. Var.}} & 
\multicolumn{1}{c}{\textbf{Coef.} \hspace{1em}} &
\multicolumn{1}{c}{\textbf{(S.E.)} \hspace{1em}} &
\multicolumn{1}{c}{\textbf{T-value} \hspace{1em}} & 
\multicolumn{1}{c}{\textbf{1-Var. Coef.}} \\
\hline
  \hline
\hline
Intercept 			& -10.027$^{***}$		 & (1.634)	 & -6.138 \\ 
\hline
\multicolumn{4}{l}{\textbf{Economic Factors}} \\
\hline
Mean GDP (\$1000s)		&   0.015$^{*}$\hspace{7pt}	&  (.008) 	&   1.901 	&  0.053$^{***}$ \\
Dif. GDP (\$1000s)		&   0.015$^{***}$		&  (.002)	&   8.000 	&  0.006$^{***}$ \\
Trade Affinity			&   0.084$^{***}$		&  (.021)	&   4.042 	&  0.354$^{***}$ \\
\hline
\multicolumn{4}{l}{\textbf{Cultural Factors}} \\
\hline
Common Civilization		&   0.663$^{***}$		&  (0.089)	&   7.441 	&  1.340$^{***}$ \\
PDI Mean			&   0.001\hspace{11pt}		&  (0.006)	&   0.122 	& -0.045$^{***}$ \\
PDI Diff. 			&   0.001\hspace{11pt}		&  (0.002)	&   0.518 	&  0.004$^{***}$ \\
IDV Mean			&   0.008\hspace{11pt}		&  (0.007)	&   1.125 	&  0.042$^{***}$ \\
IDV Diff.			&   0.014$^{***}$		&  (0.002)	&   8.337 	& -0.002\hspace{11pt} \\
MAS Mean			&  -0.003\hspace{11pt}		&  (0.006)	&  -0.406 	& -0.006\hspace{11pt} \\
MAS Diff.			&  -0.004$^{+}$\hspace{5pt}	&  (0.002)	&  -1.911 	& -0.002\hspace{11pt} \\
UAI Mean			&  -0.010$^{**}$\hspace{3pt}	&  (0.004)	&  -2.319 	& -0.006$^{***}$ \\
UAI Diff.			&  -0.010$^{***}$		&  (0.001)	&  -7.041 	& -0.002$^{***}$ \\
Gen. Trust Mean			&  -0.020$^{**}$\hspace{3pt}	&  (0.008)	&  -2.230 	&  0.038$^{***}$ \\ 
Gen. Trust Diff.		&   0.003\hspace{11pt}		&  (0.002)	&   1.253 	& -0.002\hspace{11pt} \\
Common Language			&   0.976$^{***}$		&  (.101)	&   9.585 	&   2.468$^{***}$ \\
Colonial Link			&   1.281$^{***}$		&  (.208)	&   6.162 	&   1.811$^{***}$ \\
Commonwealth Link		&   0.214\hspace{11pt}		&  (.145)	&   1.475 	&   1.540$^{***}$ \\
\hline
\multicolumn{4}{l}{\textbf{Controls}} \\
\hline
Population Avg.$^\dagger$ 	&  -0.433$^{***}$		&  (.093)	&  -4.644 	&  -0.604$^{***}$ \\	
Population Dif.$^\ddagger$ 	&  -0.024\hspace{11pt}		&  (.025)	&  -0.959 	&  -0.049\hspace{11pt} \\
Ln(Distance)			&  -0.749$^{***}$		&  (.060)	& -12.379 	&  -1.085$^{***}$ \\
Same Region			&   0.198$^{+}$\hspace{5pt}	&  (.109)	&   1.808 	&   1.849$^{***}$ \\
Contiguous Border		&  -0.253$^{**}$\hspace{3pt}	&  (.109)	&  -2.323 	&   1.721$^{***}$ \\
Visa Required			&  -0.127$^{*}$\hspace{7pt}	&  (.064)	&  -1.985 	&  -0.630$^{***}$ \\
Ln(Direct Flights + 1)		&   0.196$^{***}$		&  (.035)	&   5.557 	&   0.735$^{***}$ \\
Both in E.E.A.			&  -0.390$^{***}$		&  (.118)	&  -3.294 	&   1.287$^{***}$ \\
\hline
\multicolumn{4}{l}{\textbf{RANDOM EFFECTS}}\\
\hline \hline
&
\multicolumn{2}{c}{\textbf{Variance} \hspace{1em}} &
\multicolumn{1}{c}{\textbf{Std. dev.} \hspace{1em}}\\
\multicolumn{1}{l }{Country 1} & 
\multicolumn{2}{c}{0.182} &
0.427 \\
\multicolumn{1}{l }{Country 2} &
\multicolumn{2}{c}{0.147} &
0.384 \\
\hline
\multicolumn{1}{l }{\hspace{2em} Residual} & 
\multicolumn{2}{c}{0.512} &
0.715 \\
\hline
\end{tabular}
\small
$\dagger \ln \sqrt{\mathrm{Pop}_a \mathrm{Pop}_b}$,
$\ddagger \ln \mathrm{min}(\mathrm{Pop}_a, \mathrm{Pop}_b) / \mathrm{max}(\mathrm{Pop}_a, \mathrm{Pop}_b)$. Sample size: 1,221 relations (50 countries). Scaled deviance: 2,816. Log-likelihood: -1,491. AIC: 3,041. * p $<$ .10, ** p$<$.05, *** p$<$.01. Two-tailed tests. McFadden $R^2$: .292. One-variable model estimated using the same data as the main model, with the same random effects and a sole fixed effect for the variable of interest.
\vspace{-2em}
\end{minipage}
\end{center}
\end{table}

\setlength{\tabcolsep}{2pt}
\begin{table}[ht]
\parbox{\columnwidth}{\caption{Civilizations in LMER Model (Selected Coefs.)}}
\vspace{-1em}
\label{tbl:between-det-civ}
\begin{center}
\begin{minipage}{\columnwidth}
\begin{tabular*}{\columnwidth}{ l @{\extracolsep{3em}}  r r r}
\hline
\multicolumn{4}{l}{\textbf{FIXED EFFECTS}}\\
\hline
\multicolumn{1}{c}{\textbf{Indep. Var.}} & 
\multicolumn{1}{c}{\textbf{Coef.} \hspace{1em}} &
\multicolumn{1}{c}{\textbf{(S.E.)} \hspace{1em}} &
\multicolumn{1}{c}{\textbf{T-value} \hspace{1em}} \\
\hline
  \hline
Sinic		&  0.689\hspace{11pt} &  0.427  & 1.613 \\
Islamic        	&  1.133$^{***}$ &  0.176 &  6.428 \\
Latin American 	&  1.694$^{***}$ &  0.177 &  9.561 \\
Western        	& -0.155\hspace{11pt} &  0.142 & -1.094 \\ 
Orthodox       	&  0.878$^{**}$\hspace{4pt} &  0.425 &  2.069 \\
African        	& -0.647\hspace{11pt} &  0.444 & -1.456 \\
Buddhist       	&  1.191\hspace{11pt} &  0.753 &  1.581 \\
 \hline
\hline
\end{tabular*}
\small
Model contains same covariates as the main model in Table \ref{tbl:bet-det}, with the exception of Common Civilization. Sample size: 1,221 relations (50 countries). Scaled deviance: 2,741. Log-likelihood: -1,454. AIC: 2,977. * p $<$ .10, ** p$<$.05, *** p$<$.01. Two-tailed tests. McFadden $R^2$: .310. 
\vspace{-3em}
\end{minipage}
\end{center}
\end{table}

While it appears that Huntington's assignment of countries to civilizations is not unlike that of a community detection algorithm, the question of spuriousnes must be considered. Could it be, for instance, that Latin American countries are so strongly connected by mere accident of geographic proximity? Or could flights, colonialism, or perhaps trade flows account for the effect we witness? To try and distinguish between multiple factors influencing between-country communication we used a Linear Mixed-Effects regression to model the magnitude of edges of a nearly-complete\footnote{There were 1,221 observed counts out of a total of 1,250 possible.} weighted graph of the log transformed pairwise communication density between the 50 countries for which complete data were available for all variables of interest. By including random effects for each country, the model allows us to control for tendencies to attract more social ties due to unobserved, country-specific factors.

\textbf{Cultural Factors} Using this network of cross-country affinities, we investigate whether a battery of cultural measures covary with inter-country flows of interpersonal email communication. Following Huntington, we code countries based on location in one of the eight civilizations he demarcates, as coded by Russett, Oneal and Cox \cite{Russettetal:2000, Neumayer:2011}. A shared language should likewise increase the two countries' level of reciprocal affinity. In the very least, shared language enables communication, a logical pre-requisite for the creation of new ties between the inhabitants of two countries \cite{Neumayer:2006}. We use data regarding between-country former colonial relationships as recorded by Neumayer \cite{Neumayer:2006}, following his distinction between Commonwealth and non-Commonwealth countries. 

Additionally, we include four of Hofstede's \cite{Hofstede:1980} cultural measures: Power-Distance (PDI), Individualism (IDV), Masculinity (MAS) and Uncertainty Avoidance (UAI).\footnote{The other two Hofstede measures, Long-Term Orientation (LTO) and Indulgence vs.\ Restraint (IVR) were not included in our analysis because of insufficient coverage of the former and potential issues methodological issues created by the latter measure's comparative novelty.} To these four, we add a fifth cultural dimension, the generalized trust index, derived by Bjornskov from a meta-analysis of available studies \cite{Bjornskov:2009}. 

\textbf{Economic Factors} Our analysis is focused on two sets of economic predictors of between-country communication: development and trade. We measure economic development as the average 2011 GDP of each country pair, collected from the World Bank \cite{WorldBank:2011}. To account for the prediction of World Systems theory regarding the existence of a hierarchical structure in international relations we likewise included the absolute difference between each pair of countries GDP. Additionally, we include in our analysis a measure of bilateral trade derived from the Correlates of War Dyadic Trade dataset \cite{Barbierietal:2009, BarbieriKeshk:2012}. We defined a dyadic trade flow as the 2011 US dollar value of goods exchanged between two countries. A country's total trade was defined as the sum of a country's total imports and exports. We define the trade affinity between two countries as the natural logarithm of the ratio between the dyadic trade flow and the geometric mean between the two countries' total 
trade values. Because 49\% of all 
possible pairs of countries in our 50-country dataset had no recorded trade in the Correlates of War dataset, we used mean imputation to address the missingness issue in the trade affinity variable.\footnote{The mean imputation strategy assumes that pairs of countries for which no trade data is observed have trade affinities equal to the mean affinity recorded. Mean imputation was found to improve model fit compared to an alternative strategy of min-imputation, in which unobserved trade affinities were assumed to be equal to the minimum trade affinity recorded. Regardless of imputation strategy, the effect due to common civilization was found to persist across the models we estimated.}

\textbf{Controls} Additionally, our analysis includes several controls for factors that may systematically influence the between-country density of ties. We count here measures related to countries' populations, geographic factors related to location, and administrative factors which may impact tie formation and maintenance.

Our response variable is the between-country social affinity, defined as the ratio between the rescaled inter-country tie count, and the maximum number of possible inter-country ties. This latter quantity, defined as the product of two countries' populations is possible only in theory, however. As individuals can maintain only a limited number of social ties \cite{Dunbar:2010, Goncalves:2011}, it becomes impossible for two countries to approach this theoretical maximum. This statement is particularly important for large countries, between which densities are likely to be particularly small as a consequence of the countries' large populations. Thus, it is imperative for our model to include a control for the countries' populations.  Our measure of population uses the natural logarithm of the geometric mean of each pair of countries' populations, as derived from World Bank 2011 data\cite{WorldBank:2011}. To allow for effects due to the (potentially) peculiar nature of densities between countries of unequal 
populations (e.g.\ U.S.\ and Barbados) we also include the log-transformed ratio between the larger and the smaller country populations. 

We also control for distance, given that the density of social ties has been shown to decay exponentially with distance, a finding by Milgram \cite{TraversMilgram:1969} which has been replicated many times using offline\cite{Wellman:1979} and online data\cite{LeskovecHorvitz:2008, Takhteyevetal:2012, LibenNowelletal:2005}. Thus, the farther apart the two countries the fewer the expected ties between them, ceteris paribus. We use the log-transformed distance between each two countries' centroids, as derived by Neumayer \cite{Neumayer:2006}. Seeking to account for unevenness in the distribution of country sizes across the world, our analysis also includes a measure, collected by the Correlates of War project, of whether or not two states's territories are contiguous, either through their mainlands, or through their colonial dependencies \cite{Stinnettetal:2002}. Another factor we consider is air travel, which Takhteyev et al. (2012) found to be very strongly correlated with the geographic structure of the 
Twitter online social network\cite{Takhteyevetal:2012}. To account for the effect of air travel we use the natural log of the cumulative number of direct airline flights between each pair of countries, as recorded in the OpenFlights database\cite{OpenFlights}.

Using data collected by Neumayer \cite{Neumayer:2006}, we also measured potential administrative barriers to the creation of cross-country ties. For instance, if visa regimes make it difficult for the residents of one country to travel to another country, then one would expect fewer ties to exist between the two countries. Given the importance of European integration, we also coded countries for membership in the European Economic Area. 

\textbf{Results} We present our estimates in Table 4. The results provide support for economic as well as cultural explanations. All else being equal, wealthier countries are more likely to communicate with one another. Tie density increases by 1.5\% for each additional thousand dollars increase in a pair of countries' mean 2011 GDP per capita. As expected, inequality between two countries GDPs translates into higher communication densities, to the tune of a 1.5\% increase for each additional thousand dollars separating the two countries GDPs. For every doubling of the trade flows between two countries the model reports an increase by a factor of 1.13$(e^{.084 / \ln 2})$ in the rescaled logged communication density. 

Cultural factors also impact between-country social affinities. To test for cultural correlates of international alignment, we included four of Hofstede's \cite{Hofstede:1980} cultural dimensions: Power-Distance (PDI), Individualism (IDV), Masculinity (MAS) and Uncertainty (UAI). In addition, we included a measure indicating common membership in one of the above-mentioned civilizational blocks as a direct test of the ``clash of civilizations'' hypothesis. Common membership in the same Huntingtonian civilization nearly doubles the expected pairwise density, increasing it by a factor of $1.941$($e^{.663}$). The effects of the Hofstede measures also confirm the expected cultural homophily based on Masculinity and Uncertainty Avoidance,  but not for PDI. Each additional point difference (for variables measured on 100-point scales) yields a decrease in the rescaled tie density by 0.4\% for MAS and 1\% for UAI, while communication density decreases by 1\% for each additional point increase in the mean UAI value of 
a pair of countries. Surprisingly, cultural similarity on the IDV dimension reduces pairwise density, the opposite of what we 
expected. For each point increase in the pairwise IDV difference, we observe a 1.3\% increase in pairwise density for IDV. A shared official language has the expected strong effect, increasing pairwise tie densities by a factor of 2.70.  Additionally, Non-Commonwealth colonial relations increase communication density by a factor of 3.6 ($e^{1.281}$), while the effect of Commonwealth relations is not found to be statistically different from 0. 

With one exceptions, all control variables have significant effects on the dependent variable. The expected tie density decreases by 46\% for each doubling of the population mean. As expected, tie densities decrease drastically with distance, with a 66\% drop for each doubling of distance. Curiously, countries with contiguous borders have lower expected densities (by 22\%), \textit{ceteris paribus}. Another counter-intuitive result concerns joint membership in the European Economic Area, which is found to reduce density, by 32\%, compared to what the model would predict otherwise. Visa regimes are predicted to reduce tie density by 12\% for country pairs with unilateral or bilateral travel visa restrictions. As expected, more direct flights result in an increase of the tie density, which is predicted to increase by 33\% for every doubling of the number of direct flights between a country.

A potential issue with the results concerns the effect of model specification on the estimates. To obtain a qualitative assessment of how much our particular choice of covariates impact our findings, we estimated separate models independently for each independent variable, using the same linear mixed-effects specification and the same dataset as in the main model, but only the variable of interest as a fixed effect. With a few exceptions, our model's findings do not deviate qualitatively from the main model's estimates. While the coefficients of economic factors are robust to this comparison of model specification, there is a great deal of disagreement between the one-variable models and the full models with respect to cultural factors. The only three cultural variables where the sign, statistical significance and order of magnitude of estimates are preserved are Common Civilization, and the UAI Mean and Difference. Most controls are likewise robust to this comparison, with the exception of contiguous 
borders 
and common EEA membership, the two factors that yielded unexpected findings. Here as in the case of the non-robust Hofstede measures (all but UAI) we note the existence of interesting patterns, but we caution the reader against a decided interpretation of the main model estimates, as their signs and magnitudes appear to be sensitive to specification.  

Our analysis reveals the existence of a large, positive statistically-significant effect of common civilizational membership on between-country communication density. This result provides evidence towards a division of the world into civilizational blocks following Huntington's theory. As Table \ref{tbl:between-det-civ} reveals, not all civilizational blocks are equally consistent, however. The table shows selected estimates from a model having the same specification as the main mixed-effects model presented in Table 3, but that separates the ``common civilization'' variable according to each civilization.\footnote{Due to insufficient country-pair observations, the model could not be fit with a dummy variable for Hindu common civilization.} Three civilizations -- Latin American, Islamic and Orthodox -- have strong and significant effects when considered separately from one another. Indeed, for these civilizations the predicted effect on tie density is even higher than the overall effect shown 
in the main model. When compared against what the model would predict given two countries' values in all the other covariates, tie density is expected to increase by a factor of 2.4 for Orthodox countries, 3.1-fold for Latin American countries, and by a whopping factor of 5.44 for Latin American countries. Effects are positive but not significant for the Sinic and Buddhist civilization, possibly due to their containing few countries. Similarly insignificant are effects for the Western and African civilizations, though their coefficients are negative. 

\section{Discussion}

Not all civilizations ``survive'' a regression analysis that controls for the numerous economic and political factors that may impact cross-country communication. The strong effects we see associated with Islamic, Latin American and Orthodox countries demand further explanation however. For one reason or another, the countries in these groups have stronger level of association than the model would predict. In this respect we cautiously assign a level of validity to Huntington's contentions, with a few caveats. The first issue was already mentioned - overlap between civilizations and other factors contributing to countries' level of association. Huntington's thesis is clearly reflected in the graph presented in Figure \ref{fig:mesh}, but some of these civilizational clusters are found to be explained by other factors in Table \ref{tbl:between-det-civ}. The second limitation concerns the fact that we investigated a communication network. There is no necessary ``clash'' between countries that do not communicate,
 and Huntington's thesis was concerned primarily with ethnic conflict. Indeed, the validity of Huntington's ideas with respect to ethnic conflict has come into controversy\cite{Fox:2001}, and we limit ourselves to showing the validity -- at least partial -- of this division for communication networks. 

The third limitation is given by the data. The task of converting a worldwide communication network with uneven coverage into a set of comparable communication densities is not trivial. We hope our work on this subject, presented in Section \ref{sec:rescaling} will make a contribution to addressing this problem. We are also confident that the future growth of Computational Social Science will bring forth novel techniques for improving the estimation of such communication densities, perhaps through the incorporation of richer sets of features into the rescaling. Our experience also suggests that future studies of global online networks would benefit from an explicit consideration of the influence of market share and Internet penetration, and from the development of methods to account for potential biases due to these factors in network statistics.

Our analysis of the determinants of between-country communication likewise afforded an important opportunity to test a number of theories at the global level. The findings (unsurprisingly) support the idea that geography, transporation and administrative decisions are all important determinant of between-country communication: distance decreases density, as do visas, while direct flights increase it. Our findings in the main model with respect to contiguous borders and common European Economic Area membership appear surprising, as they decrease rather than increase density, once the other variables in the model are controlled for. These curious findings do raise the issue of potential problems with European integration, as well as of the higher potential for conflict between countries sharing borders, which may lead to less communication. We advance these explanations only tentatively however, as the direction of these coefficients appears dependent on the model specification. 

When it comes to cultural factors, it is not just Huntington's civilizations that matter. We also found important effects associated with common language, previous colonial relationships, as well as with Hofstede's uncertainty avoidance (UAI) measure. This latter finding suggests that countries with more uncertainty aversion are less likely to be connected -- perhaps because maintaining international connections requires a certain degree of risk-taking. Likewise less likely to be connected by social ties are countries that differ on this dimension, perhaps a reflection of the influence of underlying differences in social norms measured by this variable. The finding that countries that differ in the Individualism (IDV) measure are more likely to connect appears dependent on model specification, as is the result which suggests that countries with higher generalized trust are likely to have lower communication densities. We consider these findings interesting puzzles, but for which the advancement of an 
explanation is premature, given the effects' instability. As expected, we find economics to have an important role in shaping international social relations. Living in countries with higher GDP makes establishing and maintaining international connections easier, and countries with higher trade flows are also likely to have greater flows of people between them, and thus higher communication densities. We also observe an effect associated with hierarchy, as predicted by World Systems Theory: countries with dissimilar GDPs are more connected, the effect of such inequality increasing once controls are included.
\section{Conclusion}
The reality of globalization has become a commonplace of lay and scientific discourse alike. The promise of Computational Social Science is to help scholars go beyond such observations, enabling careful measurement of the world's social structures. Newly-available large, global datasets offer the possibility of an account of international relations as observed between nationals rather than among nations. Our study illustrated how such an opportunity could be pursued with one particular dataset. It is even more exciting to consider the possibility of combining insights derived from multiple online datasets to produce a clearer picture of the world's social networks. We hope our study has shown the promise of the Internet in the study of our global mesh of civilizations.


\small{
\bibliographystyle{abbrv}

}
\end{document}